\documentclass[11pt,oneside,runningheads]{taima}
\usepackage{graphicx, float, epsfig, url}
\usepackage[frenchb]{babel}
\usepackage[T1]{fontenc}
\usepackage[latin1]{inputenc}
\usepackage{graphics, enumerate}
\usepackage{amsmath,amsfonts,amstext,amssymb,color,epsfig}

\def\t{\theta}
\def\tc{\hat \theta}
\def\T{\Theta}

\def\phi{\varphi}
\def\eps{\varepsilon}

\def\N{\mathbb{N}}
\def\esp{{\rm I\mskip-4mu E}\,}

\def\MV{\text{MV}}

\def\1{ \hbox{ {\large 1} \!\!\!\!{\large I}} }
\def\qed{\hbox{$\vcenter{\vbox{
   \hrule height 0.4pt\hbox{\vrule width 0.4pt height 6pt
    \kern5pt\vrule width 0.4pt}\hrule height 0.4pt}}$}}

\begin{document}

\pagestyle{headings}
\mainmatter
\title{Codage arithmétique pour la description d'une distribution}
\pagestyle{headings}
\author{\large{Guilhem Coq\inst{1}  \and Olivier Alata \inst{2} \and Christian Olivier \inst{2} \and Marc Arnaudon \inst{1} }}
\authorrunning{G. Coq, O. Alata, C. Olivier et M. Arnaudon }
\titlerunning{Codage arithmétique pour la description d'une distribution}
\institute{
    Laboratoire de Mathématiques et Applications, UMR CNRS 6086\\
    BP 30179 - 86962 Futuroscope Chasseneuil Cedex France \\
    Tél: 05 49 49 68 97 Fax: 05 49 49 69 01\\
    \email{{coq,arnaudon}@math.univ-poitiers.fr\vspace{0.5cm}} \\
    \and
    Laboratoire Signal Image et Communication\\
    BP 30179 - 86962 Futuroscope Chasseneuil Cedex France \\
    Tél. : 05 49 49 65 67 Fax : 05 49 49 65 70 \\
    \email{alata,olivier@sic.sp2mi.univ-poitiers.fr}
}

\maketitle
\begin{abstract}
Partant du codage arithmétique prédictif adaptatif et utilisant le principe du Minimum Description Length, nous arrivons à un outil efficace pour la sélection de modèles : le critère d'information RIC. Nous présentons ensuite une extension de ces techniques de codage à l'estimation non-paramétrique d'une distribution et l'illustrons sur l'histogramme des niveaux de gris d'une image.

\vspace{0.2cm}
\textbf{Mots clés} Critères d'information, MDL, sélection de modèles, estimation non-paramétrique, histogrammes.
\end{abstract}

\section{Introduction}

Le codage arithmétique, présenté par Rissanen \cite{Rissanen_76}, est optimal en terme d'entropie. Une version simple de ce codage, pour laquelle nous renvoyons à \cite{Howard_Vitter}, est utilisée dans JPEG2000 où plusieurs modèles de référence sont utilisés. Nous présentons en partie \ref{codage} une version prédictive et adaptative, utilisée notamment dans le codeur d'images médicales CALIC, qui est un outil efficace pour la sélection de modèles. Sa longueur entre en effet dans le cadre plus général des critères d'information ou d'entropie pénalisée, introduits par exemple dans \cite{Akaike_74,Schwarz} et dont les domaines d'applications sont nombreux dès qu'il s'agit de décrire de manière optimale une distribution, citons \cite{OO,EMOJ}. Nous présentons aussi en partie \ref{histo} une procédure de sélection de la partition, non nécessairement régulière, d'un histogramme, basée sur ces techniques de codage. Elle fait suite à la méthode proposée dans \cite{Rissanen_92} et entre dans le cadre général des méthodes proposées dans \cite{Birge_06} pour la sélection d'un histogramme.

\section{Codage entropique et arithmétique prédictif adaptatif}\label{codage}

\subsection{Codage entropique}

Soit $E$ un ensemble de $m$ symboles. Un code binaire sur $E$ est une application injective $
C : E \rightarrow \cup_{i \in \N^*} \{0,1\}^i$. La longueur de $C(x)$ est notée $L(x)$. Si $L$ vérifie l'inégalité de Kraft, voir par exemple \cite{MDL}, on sait qu'elle est la longueur d'un certain code qui satisfait la condition du préfixe, indispensable au décodage. Prenant $P$ une probabilité sur $E$ et $L = \lceil - \log P \rceil$, où $\log$ est le logarithme à base 2, $L$ vérifie cette inégalité et est donc la longueur d'un code que nous confondrons avec $P$. Ainsi, si $P(x)$ est grand, $L(x)$ est faible.

Rappelons l'inégalité de convexité de Jensen : si $P$ et $Q$ sont deux probabilités sur $E$, en notant $\esp_P$ l'espérance sous $P$, on a:
\begin{equation}
\label{inégalité_information}
H(P) := \esp_P \left[ - \log P \right] \leq \esp_P \left[ - \log Q \right] =: H(P,Q)
\end{equation}

Sur des données provenant de $P$ inconnue, l'objectif est donc de trouver un codage $Q$ dont l'entropie croisée $H(P,Q)$ se rapproche de $H(P)$. A cet effet, le codage de Huffman est optimal. Cependant le codage arithmétique donne de meilleurs résultats en traitant plusieurs symboles simultanément.

\subsection{Chaînes de Markov multiples} \label{CMM}

Les Chaînes de Markov Multiples (CMM) sont le cadre naturel du codage arithmétique. Un processus $(X_n)_{n \in \N^*}$ à valeurs dans $E$ est une CMM d'ordre $k \in \N$ si $k$ est le plus petit entier vérifiant l'égalité $P(X_n|X_{n-1},\dots,X_0) = P(X_n|X_{n-1},\dots,X_{n-k})$ pour tout $n$.  Nous nous placerons toujours dans le cas où cette loi conditionnelle ne dépend pas de $n$ ; la chaîne est alors dite homogène. Une CMM d'ordre 0 est une suite de variables aléatoires indépendantes.

Prenons les $k$ premières variables d'une CMM d'ordre $k$ indépendantes et de distribution uniforme sur $E$. Notons $i \in E$ un état, $j \in E^k$ un état composé et $\t(i|j)$ la probabilité de voir apparaitre $i$ après $j$. La donnée des $(m-1)m^k$ réels $\t(i|j)$, pour $j \in E^k$ $i$ parcourant $m-1$ états de $E$, suffit à décrire l'évolution de $X$. Pour $\t$ un tel paramètre et $x^n=x_1, \dots,x_n$ une chaîne d'éléments de $E$, la vraisemblance de $x^n$ relativement à $\t$ s'écrit:
\begin{equation}
\label{vraisemblance}
P(x^n|\t) = \frac{1}{m^k} \prod_{j\in E^k} \prod_{i \in E} \t(i|j)^{n(i|j)} 
\end{equation}
avec $n(i|j)$ le nombre d'occurences de $i$ après $j$ dans $x^n$.

\subsection{Codage arithmétique prédictif adaptatif}\label{adaptatif}

Soit l'intervalle courant $I_c=[0,1[$. Soit à coder $x^n \in E^n$ à l'ordre $k$ choisi au préalable, on procède par itération. Pour $t\geq 0$ posons $x^t = x_1,\dots,x_t$ et supposons traités les $t$ premiers symboles, $t\geq 0$ ; $t=0$ signifiant que le codage n'a pas encore commencé. Pour traiter le $(t+1)-$ième, on actualise les probabilités de transitions comme suit :  
$$
\tc^{(t)}(i|j) = \frac{ n^{(t)}(i|j) + 1}{n^{(t)}(j) + m}
$$
où $i \in E$, $j \in E^k$, $n^{(t)}(i|j)$ et $n^{(t)}(j)$ sont les nombres d'occurences respectifs de $i$ après $j$ et de $j$ dans $x^t$ ; $n^{(t)}(j)$ ne devant pas compter une apparition de $j$ à la fin de cette chaîne. On pose $j=x_{t-k+1},\dots,x_{t}$ l'état actuel et on découpe $I_c$ selon les probabilités $\tc^{(t)}(i|j)$, un intervalle correspondant à un état $i$ de $E$. On choisit comme nouvel $I_c$ celui correspondant à $x_{t+1}$. 

Une fois le dernier symbole traité et notant $I_c=[a,b[$, il existe deux nombres dyadiques de longueur $\lceil - \log (b-a) \rceil$ consécutifs dans $I_c$. On prend pour code de $x^n$ la partie fractionnaire du plus grand de ces nombres. Pour illustration, prenons $E=\{a,b\}$ et codons dans la table \ref{abaa} la chaîne $abaa$ à l'ordre $k=1$. Puisque $\lceil - \log (7/24-1/4) \rceil=5$, le code 01001 convient car son prédecesseur est 01000 et $1/4 \leq 2^{-2}+2^{-5}<7/24$, $1/4 \leq 2^{-2}<7/24$.

\begin{table}
\caption{Codage arithmétique prédictif adaptatif de la chaîne $abaa$ à l'ordre $k=1$}
\centering
\begin{tabular}{cc}
\begin{tabular}{|c|c|c|c|c|}
\hline
$t$  &  $x^t$        &    $I_c$        &   $\hat \t^{(t)}(.|.)$         &     Découpage \\
\hline
0    &   $\emptyset$ &  $[0,1)$        &  $\begin{array}{c} (a|a)=1/2 \\ (a|b)=1/2 \end{array}$  &   $[0,\frac12,1)$   \\
\hline
1    &   $a$         &  $[0,\frac12)$      &  $\begin{array}{c} (a|a)=1/2 \\ (a|b)=1/2 \end{array}$  &   $[0,\frac14,\frac12)$ \\
\hline
2    &   $ab$        &  $[\frac14,\frac12)$        &  $\begin{array}{c} (a|a)=1/3 \\ (a|b)=1/2 \end{array}$  &   $[\frac14,\frac38,\frac12)$ \\
\hline
3    &   $aba$       &  $[\frac14,\frac38)$     &  $\begin{array}{c} (a|a)=1/3 \\ (a|b)=2/3 \end{array}$  &   $[\frac14,\frac{7}{24},\frac38)$ \\
\hline
4    &   $abaa$      &  $[\frac14,\frac{7}{24})$        &  $\begin{array}{c} \dots \\ \dots \end{array}$  &   \dots \\
\hline
\end{tabular}
&
\end{tabular}
\label{abaa}
\end{table}

{\bf Remarque : }lors de ce codage, nous apprenons les régularités d'ordre $k$ de la chaîne à mesure que nous la découvrons. Par conséquent, plus la chaîne est régulière à cet ordre, plus nous choisirons les grands intervalles et plus la longueur de codage sera faible.

\section{Sélection de modèles par Minimum Description Length (MDL)}

Considérons le problème de sélection de modèles suivant : étant donné une chaîne $x^n$, sélectionner l'ordre $\hat k$ d'une CMM dont $x^n$ serait une réalisation. 

Pour $k \in \N$, notons $\T_k$ le modèle des CMM d'ordre $k$ et $\T$ la réunion des $\T_k$. Le nombre de composantes libres d'un paramètre $\t \in \T_k$ est noté $|\T_k|=(m-1)m^k$. Appelons complexité stochastique de $x^n$ relativement au modèle $\T_k$ la longueur du code arithmétique de $x^n$ à l'ordre $k$, notée $C_k(x^n)$. Suivant la remarque du paragraphe précédent, si $x^n$ est une réalisation d'une CMM d'ordre $k^\star$, alors $k^\star$ minimise $C_k(x^n)$, et donc son espérance. Le MDL préconise donc de choisir pour $\hat k$ l'ordre minimisant $C_k(x^n)$ ou $\esp [C_k(x^n)]$. 

\subsection{Estimation de la complexité stochastique}

Le calcul des $C_k(x^n)$ étant complexe, Rissanen effectue dans \cite{Rissanen_86} une étude détaillée de $\esp[C_k(x^n)]$ dont le résultat essentiel est : pour $k \in \N$, $\eps > 0$, presque-tout $\t_k \in \T_k$ et $n$ assez grand on a : \begin{equation}
\label{Rissanen}
 nH(\t_k) + (1-\eps)\frac  {|\T_k|}2 \log n \leq \esp_{\t_k} \left[ C_k(x^n) \right] \leq nH(\t_k) + \frac {|\T_k|}2 \log n + o(\log n).
\end{equation}

 Il est intéressant de noter que l'inégalité de gauche est un raffinement de l'inégalité d'information de Shannon (\ref{inégalité_information}) : sur des données provenant de $\t_k$ inconnu, le codage adaptatif à l'ordre $k$ donne en moyenne un nombre de bits par symbole, $\esp_{\t_k}[C_k(x^n)]/n$, plus élevé que $H(\t_k) + |\T_k|\log n /2n$. Le terme $|\T_k|\log n /2n$ apparait ainsi comme une obstruction empêchant l'entropie du codage adaptatif de se rapprocher de l'entropie théorique $H(\t_k)$.

A partir de $x^n$, nous estimons $H(\t_k)$ par $- 1/n \log P(x^n|\tc_k)$, où $\tc_k$ est l'estimateur au sens du maximum de vraisemblance de $\t_k$ au sein du modèle $\T_k$. Les inégalités (\ref{Rissanen}) suggèrent d'estimer $\esp_{\t_k}[C_k(x^n)]$ par:
\begin{equation}
\label{RIC}
\text{RIC}(x^n,k) = - \log P(x^n|\tc_k) + \frac{|\T_k|}2 \log n,
\end{equation}
et le principe du MDL répond alors au problème de sélection de modèles posé par le choix de $\hat k = \text{Argmin} \{ \text{RIC}(x^n,k) \ | \ k \in \N \}$.

Ce critère RIC (Rissanen Information Criterion) prend la même forme que BIC (Bayesian Information Criterion) proposé par Schwarz \cite{Schwarz} et étudié dans le cadre des CMM par Zhao et al. \cite{Zhao_01}. 

On peut construire un codage arithmétique non-adaptatif, que nous appellerons simple, il est décrit dans \cite{Howard_Vitter}. Avec ce codage la chaîne $x^n$ est codée, à l'ordre $k$ et avec le paramètre $\tc_k$, en $\lceil -\log P(x^n|\tc_k)\rceil$ bits. Ainsi, minimiser la longueur de ce codage revient simplement à maximiser la vraisemblance. En termes de critères d'information, c'est donc le fait de coder de manière adaptative qui crée la pénalité $\frac{|\T_k|}{2} \log n$, permettant ainsi la sélection du bon modèle. Cela est illustré dans le paragraphe suivant.

\subsection{Comparaison des codages et critères sur simulation d'une CMM}

\begin{figure}[h]
    \centerline{\psfig{figure=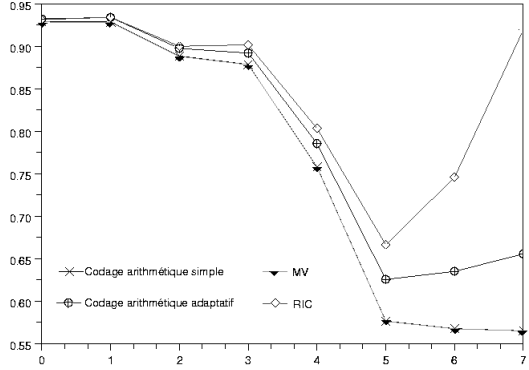,height=5cm}}
    \caption{Comparaison des différentes longueurs de codages et du critère étudié.}
    \label{fig:superposition}
\end{figure}

Nous générons une réalisation $x^n$, $n=2000$, d'une CMM d'ordre $k^\star = 5$ à 2 états. L'entropie du paramètre $\t_5$ utilisé est $H(\t_5) = 0.527$. Sur cette chaîne, pour $k=0,\dots,7$, nous effectuons le codage arithmétique simple à l'ordre $k$, le codage arithmétique adaptatif à l'ordre $k$, le calcul du maximum de vraisemblance $\MV(x^n,k)= - \log P(x^n|\tc_k)$ et le calcul de RIC$(x^n,k)$. Les résultats divisés par $n$ sont donnés en figure \ref{fig:superposition}. 

Les courbes de codage adaptatif et RIC présentent nettement un minimum en l'ordre recherché $k^\star$ ; cela s'explique par la remarque du paragraphe 2.3. L'absence d'adaptivité du codage simple justifie la superposition des deux autres courbes et donc le phénomène de surparamétrisation observé : le critère MV préfère un ordre 7.

\section{Application à la description d'une distribution par histogramme}\label{histo}

Nous présentons ici une application du codage arithmétique adaptatif au cadre non-paramétrique de l'estimation de densité par histogramme. Soit une densité $f$ inconnue sur $I$, $x^n$ un échantillon de cette distribution et $\Pi$ une partition de $I$ à $m$ intervalles $(I_j)_{j \in  [\![1,m]\!]}$. 

\subsection{Le critère proposé}

Il s'appuie sur un codage sans perte des données $x^n$, effectué  à l'aide de $\Pi$, que nous présentons ici.

Pour $i \in [\![1,n]\!]$, on note $y_i\in E=[\![1,m]\!]$ le numéro de l'intervalle dans lequel tombe $x_i$. Par indépendance des $x_i$, le codage arithmétique adaptatif de $y^n$ à l'ordre $k=0$ sera le meilleur. Notons $L(y^n|\Pi)$ la longueur de ce codage. A $i$ fixé, nous pouvons retrouver le $x_i$ correspondant à un $y_i$ en effectuant, à l'intérieur de l'intervalle $I_{y_i}$, un codage à longueur fixe $ \log l_{y_i}/r$ où $l_{y_i}$ est la longueur de $I_{y_i}$ et $r$ la précision de la machine. La longueur du code nécessaire pour retrouver $x^n$ à partir de $y^n$ est alors $L(x^n|y^n) := \sum_{j=1}^m n_j \log l_j - n \log r$ où $n_j$ est le nombre de $x_i$ tombant dans $I_j$.

La longueur du code sans perte de $x^n$ est $L(x^n|\Pi) := L(y^n|\Pi) + L(x^n|y^n)$. Il faut, pour décoder, connaitre la partition utilisée ; la longueur nécessaire à son codage étant faible devant $L(x^n|\Pi)$, nous l'omettons. Estimons $L(y^n|\Pi)$ par RIC$(y^n,0)$ (\ref{RIC}) et définissons le nouveau critère:
$$
\text{Crit}(x^n|\Pi) = \text{RIC}(y^n,0) + L(x^n|y^n) = -\log P(y^n|\tc_0) + \frac{m-1}{2} \log n + \sum_{j=1}^m n_j \log l_j - n \log r
$$
où $\tc_0(j) = n_j/n$ est estimé au sens du maximum de vraisemblance. Utilisant (\ref{vraisemblance}) il vient
\begin{equation}
\label{critere}
\text{Crit }(x^n|\Pi) = - \sum_{j=1}^m n_j \log \frac {n_j}{nl_j} + \frac{m-1}{2} \log n -n \log r
\end{equation}
qui entre dans le cadre général des critères utilisés par exemple par Birgé \cite{Birge_06} pour la sélection d'un histogramme.

Le principe du MDL préconise de choisir pour partition celle qui minimise ce critère. Le nombre de partitions de $I$ étant trop élevé, on se restreint à la classe des sous-partitions d'une partition $\Pi_{\hbox{max}}$ à $R$ intervalles donnée. Nous utilisons la méthode de programmation dynamique proposée par Rissanen et al. \cite{Rissanen_92} qui permet de trouver la sous-partition optimale de $\Pi_{\hbox{max}}$ en $cR^2$ opérations seulement, où $c$ est une constante.

\subsection{Exemples}
\begin{figure}
\begin{center}
  \begin{tabular}{ccc}
    \psfig{figure=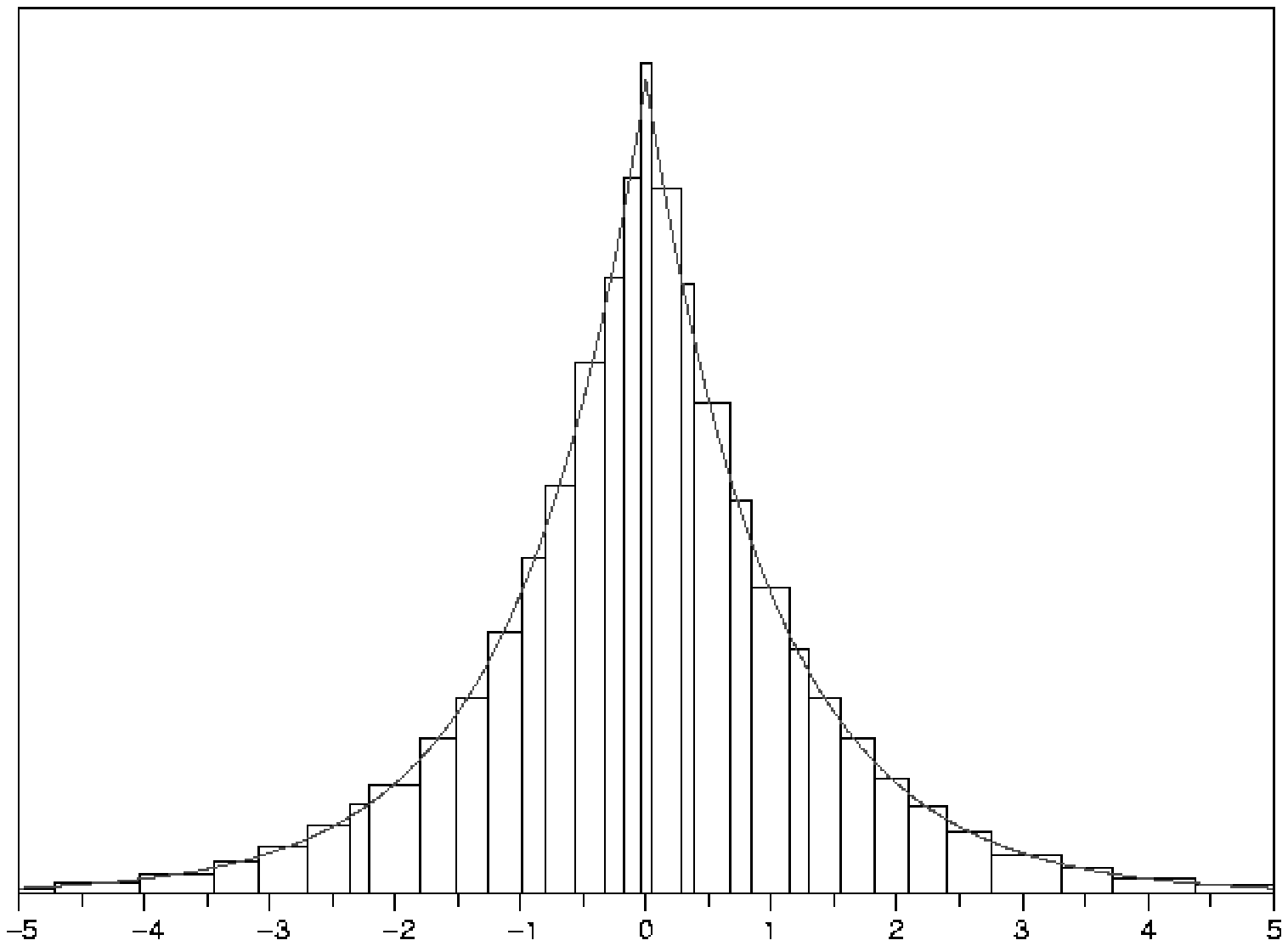,width=4.8cm} & 
    \psfig{figure=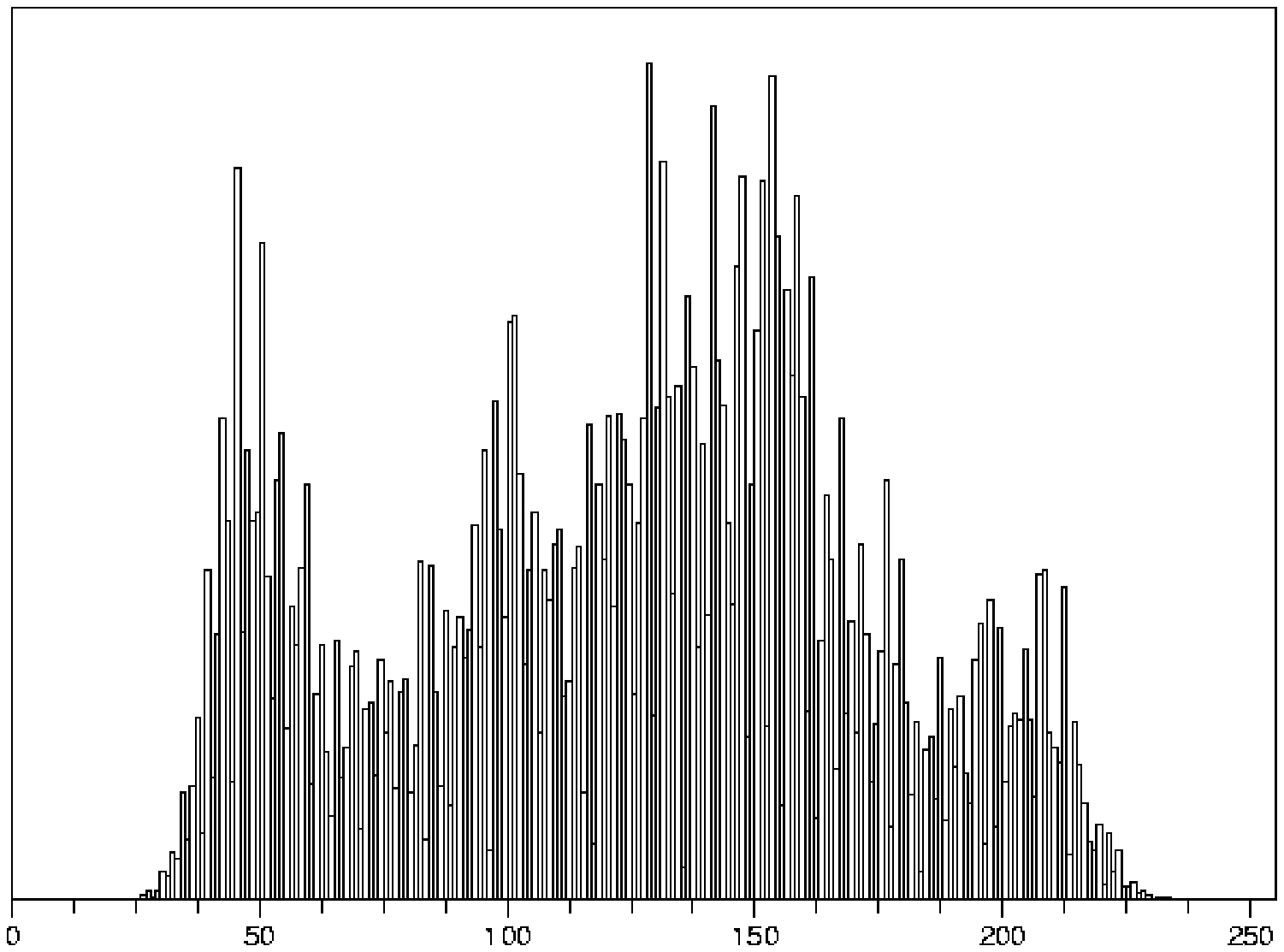,width=4.8cm} & 
    \psfig{figure=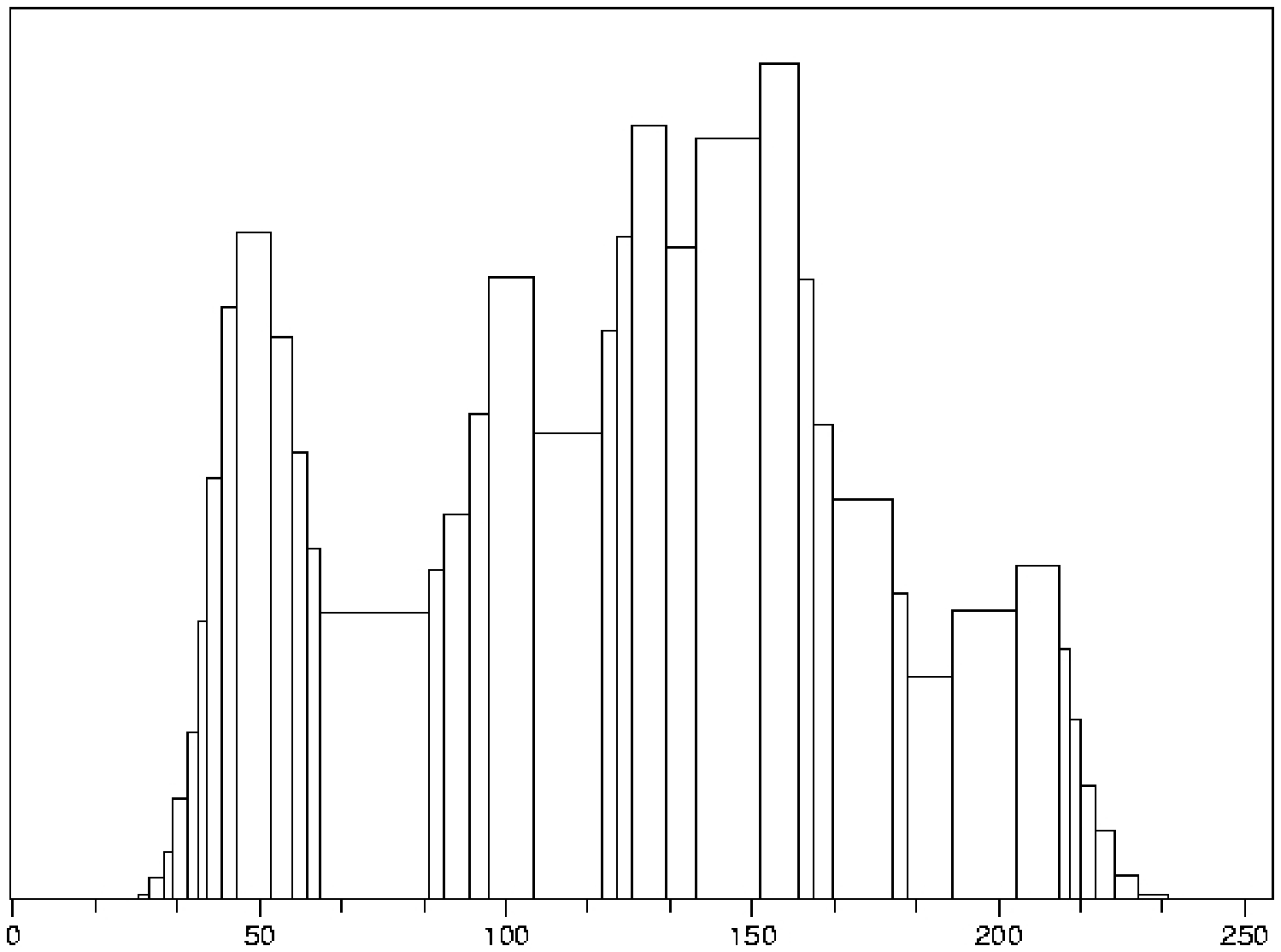,width=4.8cm} 
    \\
    (a) & (b) & (c)
  \end{tabular}

    \caption{Histogramme choisi sur une Laplacienne (a) et histogramme choisi sur l'image Léna en 39 classes (c) à partir de la distribution initiale en 256 classes (b).}
    \label{fig:histos}
\end{center}
\end{figure}

\begin{figure}
  
  \begin{center}
  \begin{tabular}{cc}
    \psfig{figure=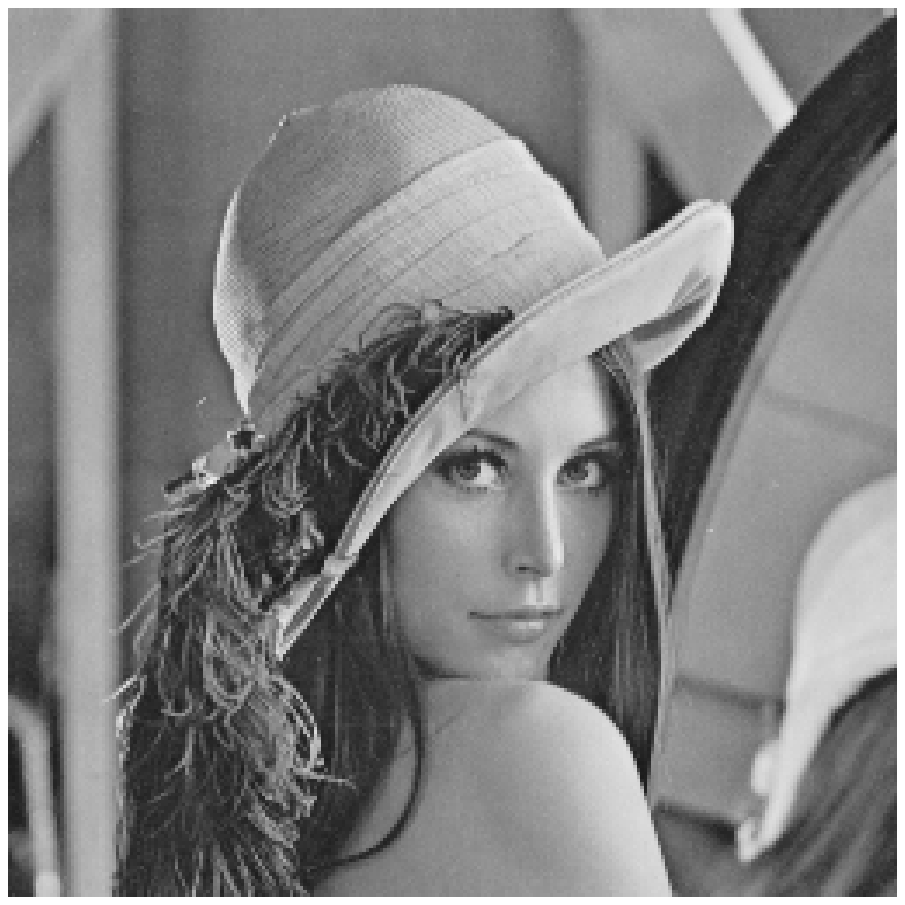,width=4.8cm} & 
    \psfig{figure=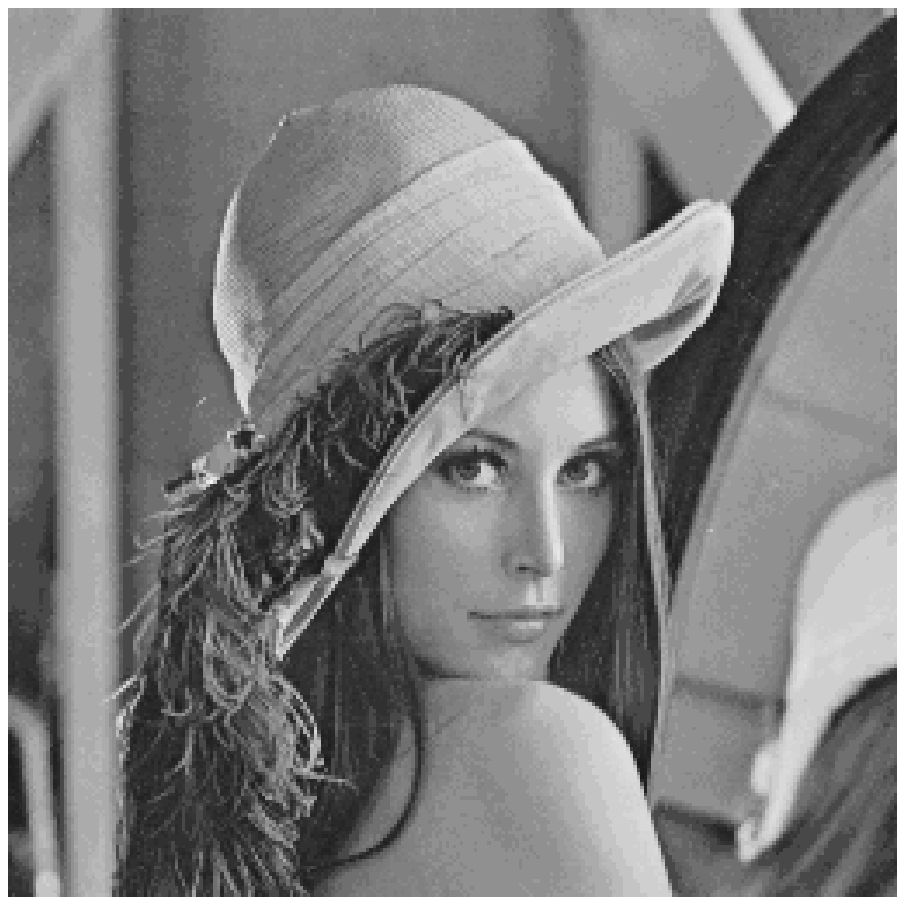,width=4.8cm}
    \\
    (a) & (b)
  \end{tabular}
  \end{center}
      
   \caption{L'image originale (a) et l'image reconstruite (b) sur 39 niveaux pour un PSNR de 38,52 dB.}
   \label{fig:lena}

\end{figure}

Pour une densité Laplacienne $e^{-|x|}/2$ (par exemple une distribution de coefficients AC de blocs DCT $8 \times 8$ dans JPEG), avec $I=[-5,5]$ et $\Pi_{max}$ la partition régulière de pas $2.10^{-2}$, on obtient la partition présentée en figure \ref{fig:histos}.(a) ; la distribution théorique est également représentée. Sur l'histogramme des 256 niveaux de gris de l'image Léna, avec $I=[0,255]$, la partition choisie \ref{fig:histos}.(b) a 39 intervalles.  Dans les deux cas, le critère choisit plus d'intervalles aux endroits où la densité présente de fortes variations. La reconstruction de l'image Léna sur les 39 niveaux de gris choisis est donnée en figure~\ref{fig:lena}.(b). Elle présente un PSNR de 38,52 dB par rapport à l'image originale et est visuellement très acceptable.


\subsection{Conclusion}

Nous avons présenté, à partir du MDL, un procédé de description d'une distribution par un histogramme. L'obtention d'un tel histogramme à partir des données numériques peut-être exploitée dans un contexte de reconnaissance de formes. De plus, l'utilisation du critère présenté peut aussi être d'un intérêt certain dans la chaîne de codage source d'une image ou d'une vidéo.

\end{document}